\documentclass{article}
\usepackage{amssymb}
\usepackage{spconf,amsmath,graphicx}
\usepackage{cleveref}  
\usepackage{url}
\usepackage{longtable}
\usepackage{array}
\usepackage{ragged2e}
\usepackage{lscape}
\usepackage{booktabs}
\usepackage{makecell}
\usepackage{varwidth}

\title{GENERATIVE MODELS FOR IMPROVED NATURALNESS, INTELLIGIBILITY, AND VOICING OF WHISPERED SPEECH}
\name{Dominik Wagner$^1$, Sebastian P. Bayerl$^1$, H\'ector A. Cordourier Maruri$^2$, Tobias Bocklet$^{1,2}$}
\address{
 $^1$Technische Hochschule Nürnberg Georg Simon Ohm, Germany\\
 $^2$Intel Labs
 }

\newcommand{\cf}{\mbox{cf.\ }}
\newcommand{\ie}{\mbox{i.\,e.}}

\newcommand{\etal}{\mbox{\emph{et.\,al.\ }}}

\makeatletter
\def\ps@IEEEtitlepagestyle{%
\def\@oddfoot{\mycopyrightnotice}%
\def\@evenfoot{}%
}

\copyrightnotice{978-1-6654-7189-3/22/\$31.00~\copyright2023 IEEE}
\begin{document}

\maketitle
\begin{abstract}
This work adapts two recent architectures of generative models and evaluates their effectiveness for the conversion of whispered speech to normal speech. We incorporate the normal target speech into the training criterion of vector-quantized variational autoencoders (VQ-VAEs) and MelGANs, thereby conditioning the systems to recover voiced speech from whispered inputs. Objective and subjective quality measures indicate that both VQ-VAEs and MelGANs can be modified to perform the conversion task. We find that the proposed approaches significantly improve the Mel cepstral distortion (MCD) metric by at least 25\% relative to a DiscoGAN baseline. 
Subjective listening tests suggest that the MelGAN-based system significantly improves naturalness, intelligibility, and voicing compared to the whispered input speech. 
A novel evaluation measure based on differences between latent speech representations also indicates that our MelGAN-based approach yields improvements relative to the baseline.
\end{abstract}
\vspace{-1mm}
\section{Introduction}
\label{sec:intro}
\vspace{-1mm}
The melody of an utterance is communicated mainly by alternating the pitch of one's voice over time \cite{laver94phonetics}. 
Humans are sensitive to these changes in fundamental frequency ($F0$) and can easily recognize its absence \cite{hart90perceptual}.
An important characteristic of whispering is the absence of pitch, which is the main reason why whispered speech is perceived as less natural sounding than normal speech \cite{vanas01tracheo}. 
Nevertheless, whispered speech is often used to avoid disturbing others in quiet places such as libraries and to avoid being overheard by unwanted listeners. 
The estimation of natural sounding speech from whispered speech signals remains a challenging task, despite the considerable progress in generative modeling and neural speech synthesis for regular utterances. 
The reconstructed signals often contain artifacts and lack the expressiveness of normal speech. 

Previous works use Gaussian mixture models \cite{toda05whspgmm}, transformer networks \cite{niranjan2021endtoend}, bidirectional long short-term memory networks \cite{meenakshi18whsp}, and other sequence-to-sequence models \cite{lian19whisper} for the conversion task. 
Several other studies employ generative adversarial networks (GANs) \cite{goodfellow14gan}. 
In \cite{teng21whsp}, an attention component is added to a GAN model, in order to achieve implicit time alignments between whispered and normal speech. 
In \cite{pascual18whisper}, a speaker-dependent speech enhancement GAN \cite{pascual17segan} is adapted for whisper to normal speech conversion. 
The work by Parmar \etal \cite{parmar19whisper} compares two GAN architectures (DiscoGAN \cite{kim17discogan} and CycleGAN \cite{zhu17cyclegan}) for whisper to normal speech conversion. 
Their approach relies on separate systems for the prediction of $F0$ and cepstral features, as well a statistical vocoder for speech synthesis \cite{erro14aho}. 

In contrast to previous works, we avoid estimating the fundamental frequency of the normal speech with a separate model and train speaker-independent MelGANs \cite{kundan19melgan} and vector-quantized variational autoencoders (VQ-VAEs) \cite{vandenoord17vqvae}. 
Our models are either capable of directly performing the conversion between waveforms, or can be transformed into an end-to-end system. 
We chose MelGAN for our experiments, because it achieved promising results in regular end-to-end speech generation tasks \cite{kundan19melgan}.
VQ-VAEs were chosen due to their fundamentally different approach, which has also been shown to be capable of accurately reconstructing spectrogram features \cite{chorowski19unsupervised}. 

In this work, we incorporate differences in the characteristics of the input signal and the desired output signal into the training criterion of MelGAN and VQ-VAE models. 
We then determine the objective and subjective quality improvements and benchmark our systems against a DiscoGAN model. 
Our contributions are:\footnote{Audio samples are available at:\\ \scriptsize{\url{https://th-nuernberg.github.io/speech-conversion-demo}}}
\vspace{-0.5mm}
\begin{itemize}
    \item Introduction of two new speaker-independent whisper to normal speech conversion (SC) systems called \textit{SC-MelGAN} and \textit{SC-VQ-VAE}
     \vspace{-2mm}
    \item Comparison of state-of-the-art methods for speech conversion on the wTIMIT corpus \cite{lim10wtimit} 
     \vspace{-2mm}
    \item Transformation of the VQ-VAE model into a vocoder-free end-to-end conversion system
    \vspace{-2mm}
    \item Introduction of a novel method for objective quality measurement based on latent speech representations
\end{itemize}
\section{Method}
\label{sec:method}
\vspace{-1mm}
MelGAN and VQ-VAE models were designed to generate human speech that resembles the characteristics of the provided input as closely as possible. 
However, our goal of converting whispered speech to normal speech is a task, where the characteristics of the provided inputs are different to those of the desired outputs. 
Therefore, we modify the models with regard to their objective function. 
Additional alterations to the VQ-VAE architecture are made to enable end-to-end speech conversion.
\vspace{-2mm}
\subsection{Alignment}
In order to equalize the sequence lengths, we use dynamic time warping (DTW) on the whispered and normal pairs of data. 
First, we apply volume normalization and trimming of leading and trailing silence to each audio sample. 
Silent frames are detected with an off-the-shelf voice activity detector. 
The trimming procedure reduced the average length of whispered signals by approximately 3 seconds and the length of normal signals by approximately 1.5 seconds, and proved to be crucial to obtain alignments of sufficient quality. 
Second, we compute mel-spectrogram features for each volume-normalized and trimmed audio sample with window length of 1024, hop length 256 and 80 Mel channels at a sample rate of 22.05 kHz. 
The same parameters are also used to compute input features for the proposed models. 
We experimented with varying hop lengths and window sizes but found the impact on the overall alignment quality negligible. 
Third, the extracted whispered and normal mel-spectrogram features are passed to the DTW algorithm, to compute the optimal alignment path in the frequency domain using $\ell_2$-distance as the cost function. 
Since the alignment path needs to be applied in the time domain, we use linear interpolation to bootstrap the path to the resolution of the raw audio signal. 
Finally, the indices along the bootstrapped time domain alignment are used to equalize the lengths of each signal pair.
\vspace{-2mm}
\subsection{SC-MelGAN}\label{ssec:melgan_mod}
MelGAN \cite{kundan19melgan} is a fully convolutional feedforward architecture for audio waveform synthesis. 
Its generator component employs stacks of transposed convolutional layers, which are used to upsample the input mel-spectrogram until the same temporal resolution as the target waveform is reached.
The kernel size in each upsampling layer is determined as a multiple of stride (i.e. the upsampling factor). 
Upsampling is done in four stages with strides of 8, 8, 2 and 2, which corresponds to an overall upsampling factor of 256. 

Three residual blocks with dilated convolutional layers are inserted after each upsampling layer. 
Dilation grows as a power of the kernel size. 
The kernel size is 3, which leads to dilation factors of 1, 3, and 9 in each residual block. 
Additionally, each layer of the generator is weight-normalized.
Except for the tanh activation function at the output layer, leaky-relu is used for activation throughout the generator. 

The discriminator is comprised of three blocks with identical structure, each receiving differently scaled versions of the audio. 
The first discriminator block $D_1$ acts on the scale of the final raw audio, whereas $D_2$ and $D_3$ act on audio that is downsampled by factors of 2 and 4 respectively. 
Downsampling is achieved via average pooling with kernel size 4. 
Each discriminator block consists of a convolutional input layer (kernel size 15, stride 1), four intermediate convolutional layers (kernel size 41, stride 4), and two final convolutional layers (kernel sizes 5 and 3). 
All discriminator components use leaky-relu for activation. 

The following equations describe the necessary adjustments to the training objective of the original MelGAN system, which enforce the reconstruction of voiced speech from a given whispered utterance. 
The subscript $norm$ refers to normal speech and $whsp$ refers to whispered speech.
The discriminator objective introduced in \cite{kundan19melgan} is redefined as follows:
\begin{equation}\label{eq:melgan_d_loss_mod}
\begin{split}
\underset{D_k}{\mathrm{min}} & 
\mathbb{E}_{\mathbf{x}} \left[ \mathrm{max} (0, 1 - D_k(\mathbf{x}_{norm})) \right] \\
& + \mathbb{E}_{\mathbf{s}} \left[ \mathrm{max} (0, 1 + D_k(G(\mathbf{s}_{whsp}))) \right], \, \forall k,
\end{split}
\end{equation}
where $\mathbf{s}_{whsp}$ is the mel-spectrogram of the whispered speech and $\mathbf{x}_{norm}$ represents the raw audio waveform of the normal speech recordings. 
The normal training target waveform $\mathbf{x}_{norm}$ is passed through each of the $k$ discriminator components.
The conditioning information of whispered speech $\mathbf{s}_{whsp}$ is passed through the generator $G$ and subsequently processed by each of the discriminator components $D_k$. 
The generator receives mel-spectrogram information from whispered speech:
\begin{equation}\label{eq:melgan_gen_mod}
\underset{G}{\mathrm{min}} \:\: \mathbb{E}_{\mathbf{s}} \left[ \sum_{k=1}^{N} -D_k (G(\mathbf{s}_{whsp})) \right].
\end{equation}
The cost of feature matching is represented by the $\ell_1$-loss between the discriminator outputs of the normal audio $\mathbf{x}_{norm}$ and the waveform generated from whispered mel-spectrograms $\mathbf{s}_{whsp}$:
\begin{equation}\label{eq:melgan_fm_mod}
\mathbb{E}_{\mathbf{x}, \mathbf{s}} \left[  \sum_ {i=1}^{T} \sum_{k=1}^N || D_k^{(i)}(\mathbf{x}_{norm}) - D_k^{(i)}(G(\mathbf{s}_{whsp})) ||_1  \right].
\end{equation}
$T$ is the number of layers in each discriminator block and $N$ denotes the number of discriminator blocks. 
\vspace{-2mm}
\subsection{SC-VQ-VAE}\label{ssec:own_vqvae}

Variational autoencoders (VAEs) \cite{kingma13vae} learn mappings between the observed data $\mathbf{x} \in \mathcal{X}$ and a $k$-dimensional latent representation $\mathbf{z} \in \mathbb{R}^k$ using the autoencoder framework. 
The decoder $p_{\theta}(\mathbf{x} | \mathbf{z})$ with the parameter set $\theta$ generates the data $\mathbf{x}$ from a latent representation $\mathbf{z}$, which is sampled from a prior distribution $p(\mathbf{z})$.
The encoder $q_{\phi}(\mathbf{z} | \mathbf{x})$ with parameters $\phi$ approximates the true but intractable posterior $p( \mathbf{z} | \mathbf{x} )$. 
Encoder and decoder are jointly trained to maximize the lower bound on the log-likelihood of the data $\mathbf{x}$. 

Vector-quantized variational autoencoders (VQ-VAEs) \cite{vandenoord17vqvae} replace continuous latent representations with discrete ones using vector quantization. 
VQ-VAEs define a codebook consisting of $K$ embedding vectors $\mathbf{e}_i \in \mathbb{R}^d, \, i \in 1, 2, \ldots , K$, where $d$ is the dimensionality of each embedding vector. 
The input $\mathbf{x}$ is passed through the encoder network producing the output $z_{e}(\mathbf{x})$ prior to quantization. 
Representations generated by the encoder are then replaced by an embedding vector using a nearest neighbor lookup. 
The closest embedding vector $\mathbf{e}_n$ is passed to the decoder as the latent representation $\mathbf{z}$. 
The VQ-VAE loss function consists of three terms:
\begin{equation}\label{eq:vqvaeloss}
\begin{split}
\mathcal{L}_{vqvae} = log \, p(\mathbf{x} | \mathbf{z} = \mathbf{e}_n) + \\ 
                || sg \left[z_e(\mathbf{x}) \right] - \mathbf{e}||^2_2 + \\
                \gamma || z_e(\mathbf{x}) - sg \left[\mathbf{e} \right] ||^2_2. 
\end{split}
\end{equation}
The first term is the reconstruction loss, which optimizes encoder and decoder via straight-through gradient estimation \cite{bengio13sg}.
The second term trains the embedding space by moving each embedding vector $\mathbf{e}_i$ towards the encoder outputs $z_e(\mathbf{x})$. 
$sg[\cdot]$ denotes the stop gradient operation, which indicates that its operand remains constant throughout training. 
The third term ensures that the encoder commits to an embedding, with $\gamma$ being a hyperparameter for scaling. 
As suggested in \cite{vandenoord17vqvae}, we use $\gamma = 0.25$ in our experiments. 

Our \textit{SC-VQ-VAE} encoder consists of five convolutional layers (kernel sizes 3, 3, 4, 3, and 3) followed by three residual blocks. 
The input signal is downsampled by a factor of 2 at the third convolutional layer. 
Each residual block consists of two convolutional layers with kernel sizes 3 and 1. 
The encoder output is passed to the vector quantizer, to determine the embedding vector. 
The decoder consists of a convolutional layer with kernel size 3, the same residual stack that is used for the encoder, and three transposed convolutional layers with kernel size 3. 
The decoder generates mel-spectrogram features from the embedding vector, which are subsequently synthesized with WaveGlow \cite{prenger18waveglow}.
An alternative decoder implementation employs the generator component from MelGAN and directly produces the speech waveform. 
This enables us to train the system in an end-to-end fashion. 
The VQ-VAE training objective is changed with respect to the reconstruction loss.
We determine the reconstruction quality as the $\ell_1$-error between the mel-spectrogram of the converted whispered input utterance and the mel-spectrogram of the corresponding normal target utterance.  
The overall loss is weighted equally between the reconstruction loss and the vector quantization loss. 

To mitigate codebook collapse \cite{vandenoord18collapse}, we initialize the codebook by applying $K$-means clustering to the encoder outputs, thereby ensuring that all embedding vectors are initially likely to be used. 
We use $K=256$, where 256 is the number of embedding vectors.  
Similar approaches to avoid codebook collapse have been previously proposed in \cite{roy18vqvaeinit,lancucki20robust}. 
\vspace{-8mm}
\subsection{Quality Assessment}
Root mean squared error (RMSE), Mel cepstral distortion (MCD) and Pearson correlation were chosen for objective quality assessment. 
Three comparative mean opinion score (CMOS) tests were conducted to determine the subjective quality of the proposed systems. 
Additionally, we propose a new evaluation method based on pairwise distances of latent speech representations. 

The RMSE between normal and converted speech signals was computed similarly to \cite{parmar19whisper,taal10intelligiblity}, after aligning all frames in the utterance via DTW. 
MCD refers to the sum of squared distances between the normal and the converted cepstral coefficients summed over all frames in the signal \cite{mashimo01eurospeech}. 
Following \cite{taal10intelligiblity}, we also report Pearson's correlation coefficient to determine the strength of the relationship between the converted fundamental frequency ($F0$) features and the normal $F0$ features. 
The cepstral and $F0$ features were extracted with Ahocoder \cite{erro14aho}. 

We evaluated the subjective quality in terms of naturalness, intelligibility, and voicing in three CMOS tests. 
Two audio samples were played in each trial.
Sample A represented a transformation of the whispered sample B.
The participants were asked to rate, whether they find the transformed version (sample A) more voiced, more natural, and more intelligible than the whispered version (sample B).
Following \cite{dinh20alaryng}, responses were selected from a 5-point scale that consisted of the choices ``definitely better'' (+2), ``better'' (+1), ``same'' (0), ``worse'' (-1), and ``definitely worse'' (-2). 
Additionally, the listeners were asked, whether they would prefer a dialog partner who talks in the voice of sample A or in the voice of sample B. 
All 16 participants listened to 23 randomly selected samples from the test set. 
The utterances remained the same across all 4 model variants. 
Consequently, each listener had to listen to a total of $4 \times 23 = 92$ converted utterances and their original whispered counterparts. 
The first 3 samples of each model variant served as training samples to introduce the listeners to the task. 
Hence, $4 \times 3 = 12$ per model were excluded from the final evaluation. 
The remaining 20 samples for each model ($4 \times 20 = 80$ samples in total) were used to compute the CMOS scores shown in \Cref{tab:cmos}. 

Hidden representations obtained from wav2vec 2.0 \cite{baevski20w2v2} (W2V2) models have been successfully employed in phoneme recognition, speech emotion recognition, dysfluency detection, and vocal fatigue detection \cite{baevski20w2v2,pepino21_interspeech,bayerl22ksof,vocfatigue22}. 
These works demonstrate that W2V2 embeddings are well suited to encode speaker and language characteristics, as well as characteristics of speech disorders. 

We used a wav2vec 2.0 model pretrained and finetuned on 960 hours of LibriSpeech data to extract 768-dimensional embeddings at the second of its 12 transformer blocks. 
We computed the pairwise cosine distance between embeddings representing the converted utterances and their underlying whispered and normal versions. 
The reconstruction quality is assumed to be better, the smaller the distance between an embedding obtained from a converted utterance and its corresponding normal embedding. 
\vspace{-2mm}
\section{Experiments}
\label{sec:experiments}
\vspace{-1mm}
We investigated the whispered to normal speech conversion capabilities of four speaker-independent models. 
\textit{DiscoGAN} represents the system introduced in \cite{parmar19whisper} and serves as the baseline. 
However, we used all speakers in the wTIMIT corpus (except for the ten speakers in the test set) as training data for better comparison with our other speaker-independent systems. 
\textit{SC-MelGAN} is the adapted MelGAN architecture described in \Cref{ssec:melgan_mod}. 
We implemented two variants of the modified VQ-VAE model introduced in \Cref{ssec:own_vqvae}. 
\textit{SC-VQ-VAE+GAN} employs the generator from MelGAN as its decoder component. 
The generator directly produces the speech waveform and a vocoder is not required. 
\textit{SC-VQ-VAE+WG} uses a decoder comprised of convolutional layers and WaveGlow \cite{prenger18waveglow} for speech synthesis. 
\vspace{-2mm}
\subsection{Data}
We used the whispered TIMIT (wTIMIT) corpus \cite{lim10wtimit} for our experiments.
Each speaker in the wTIMIT corpus utters 450 phonetically compact sentences from the TIMIT \cite{garolfo93timit} database in both a normal and a whispered manner. 
The corpus consists of 24 female speakers and 25 male speakers from two main accent groups (Singaporean-English and North-American English).
We excluded ten speakers (five female and five male) from the training data, to serve as the test set for quality assessment. 
\vspace{-2mm}
\subsection{Objective Evaluation}\label{ssec:obj}
\Cref{tab:measures_all_spk} shows RMSE and Pearson correlation coefficient w.r.t. normal and converted $F0$ sequences, as well MCD (mean and standard deviation) over all utterances in the test set. 
All models were capable of generating lower RMSE scores than the whispered reference signals. 
Except for \textit{DiscoGAN}, the approaches also achieved better MCD results than the whispered speech. 
The two \textit{SC-VQ-VAE} systems yielded the lowest average cepstral distortion scores with improvements of 34\% (\textit{SC-VQ-VAE+GAN}) and 27\% (\textit{SC-VQ-VAE+WG}) relative to the DiscoGAN baseline. 
The strongest linear relation between normal and converted $F0$ sequences was produced by the \textit{SC-MelGAN} system.
\textit{SC-MelGAN} also showed the strongest RMSE improvement (25\%) compared to the baseline.
\setlength{\tabcolsep}{7pt} 
\renewcommand{\arraystretch}{1} 
\begin{table}
\caption{RMSE and correlation of pitch sequences, as well as MCD results for the different whisper to normal speech conversion models. MCD results for \textit{SC-MelGAN}, \textit{SC-VQ-VAE+GAN}, and \textit{SC-VQ-VAE+WG} show significant improvements relative to the whispered reference ($p<.001$, two-sample two-tailed t-test).}
\label{tab:measures_all_spk}
\centering
\scalebox{0.95}{\begin{tabular}{lccccc}
\toprule
      \textbf{Experiment} &  \textbf{RMSE}  & \textbf{Corr} &  \multicolumn{2}{c}{\textbf{MCD}}  \\
                         &                 & &  \footnotesize{mean} & \footnotesize{std}  \\
\midrule
         DiscoGAN      & 16.10 & 0.62 & 9.77  & 1.39 \\
         SC-MelGAN    & {\bf 12.16}  & {\bf 0.64} & 7.35  & 0.47 \\
      SC-VQ-VAE+GAN     & 14.69 & 0.42 & {\bf 6.50} & 0.51  \\
      SC-VQ-VAE+WG     & 15.24 & 0.37  & 7.07 & 0.39   \\
       \midrule
       \emph{Reference}  & \emph{16.31} & \emph{0.08}&  \emph{9.75} & \emph{1.19}\\
\bottomrule
\end{tabular}}
\end{table}
\vspace{-1mm}
\begin{figure}[!htb]
\vspace{-4.5mm}
  \centering
	\includegraphics[width=0.48\textwidth]{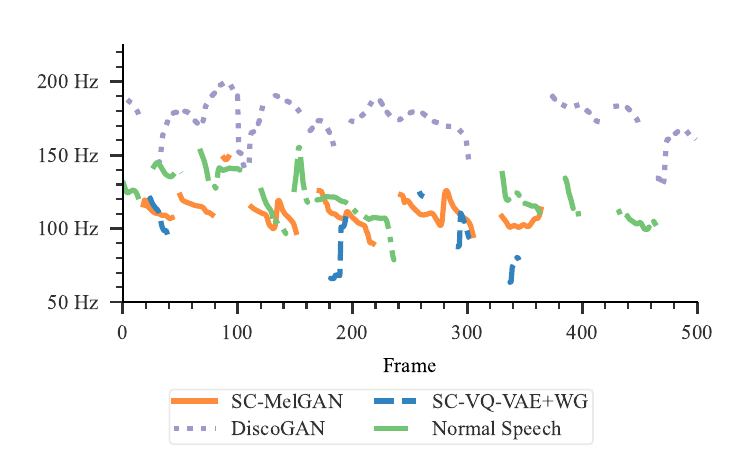}
	\vspace{-8.5mm}
    \caption{Pitch contour comparison for the utterance ``It's impossible to deal with bureaucracy.''.}
  \label{fig:pitch}
\vspace{-5mm}
\end{figure}

\Cref{fig:pitch} visualizes the pitch contours of the utterance ``It's impossible to deal with bureaucracy.'' spoken by male speaker \emph{014} from the test set. 
In this example, the fundamental frequency of \textit{DiscoGAN} (dotted line) lies above the normal speech target (solid green line) throughout the utterance. This leads to a converted voice that sounds more high-pitched than its normal counterpart and contributes to the comparatively high RMSE score of the \textit{DiscoGAN} model (\cf \Cref{tab:measures_all_spk}). 
Our \textit{SC-MelGAN} system (solid orange line) produced a pitch contour in a frequency range close to normal speech (solid green line). 
The \textit{SC-VQ-VAE+WG} model (dashed blue line) generated several voiced frames with frequencies located below those of the normally spoken utterance, while the \textit{SC-VQ-VAE+GAN} did not yield any voiced frames in this example and is thus not shown in \Cref{fig:pitch}.
\vspace{-2mm}
\subsection{Subjective Evaluation}\label{ssec:subj}
\Cref{{tab:cmos}} shows the comparative mean opinion scores of whispered and converted sample pairs for intelligibility, naturalness, and voicing. 
Positive scores indicate improvement over whispered speech. 
Our \textit{SC-MelGAN} approach achieved improvements across all three CMOS tests. 
The most noticeable improvement yielded the test for voicing with a score of 1.11. 
The improvements obtained with the \textit{SC-MelGAN} system were statistically significant ($p<.001$) for intelligibility, naturalness, and voicing as compared to zero, \ie, no preference, in a one-sided one-sample t-test.
The baseline system (\textit{DiscoGAN}) was also able to significantly ($p < .001$) improve voicing with a score of 0.62, but intelligibility and naturalness were perceived as worse with scores of -0.36 and -0.53. 
Our two models based on variational autoencoders (\textit{SC-VQ-VAE+GAN} and \textit{SC-VQ-VAE+WG}) yielded worse subjective evaluation results than the \textit{DiscoGAN} baseline. 
\setlength{\tabcolsep}{5pt} 
\renewcommand{\arraystretch}{1} 
\begin{table}
\caption{CMOS test results comparing the utterances converted by each system against the whispered input speech.}
\label{tab:cmos}
\centering
\scalebox{0.95}{\begin{tabular}{lccc}
\toprule
      \textbf{Experiment} &  \textbf{Intelligibility}  & \textbf{Naturalness} &  \textbf{Voicing}  \\
\midrule
         DiscoGAN            & -0.36                   & -0.53                   & 0.62 \\
         SC-MelGAN           & {\bf 0.13}              & {\bf 0.25}              & {\bf 1.11}  \\
      SC-VQ-VAE+GAN          & -1.21                   & -1.62                 & -0.34 \\
      SC-VQ-VAE+WG           & -0.99                   & -1.14                   & -0.03  \\
\bottomrule
\end{tabular}}
\vspace{-5mm}
\end{table}

When asked, which sample they would prefer in a conversation, the listeners preferred the sample converted with our \textit{SC-MelGAN} system over the original whispered sample in 62\% of the trials. 
The \textit{DiscoGAN} baseline was preferred over the whispered sample in 25\% of the trials.
Our two VAE-based systems were preferred in 8\% (\textit{SC-VQ-VAE+WG}) and 2\% (\textit{SC-VQ-VAE+GAN}) of the trials. 
\vspace{-2mm}
\subsection{Wav2vec 2.0 Embeddings}
\vspace{-1mm}

\begin{figure}[!htb]
\vspace{-2mm}
  \centering
	\includegraphics[width=0.48\textwidth]{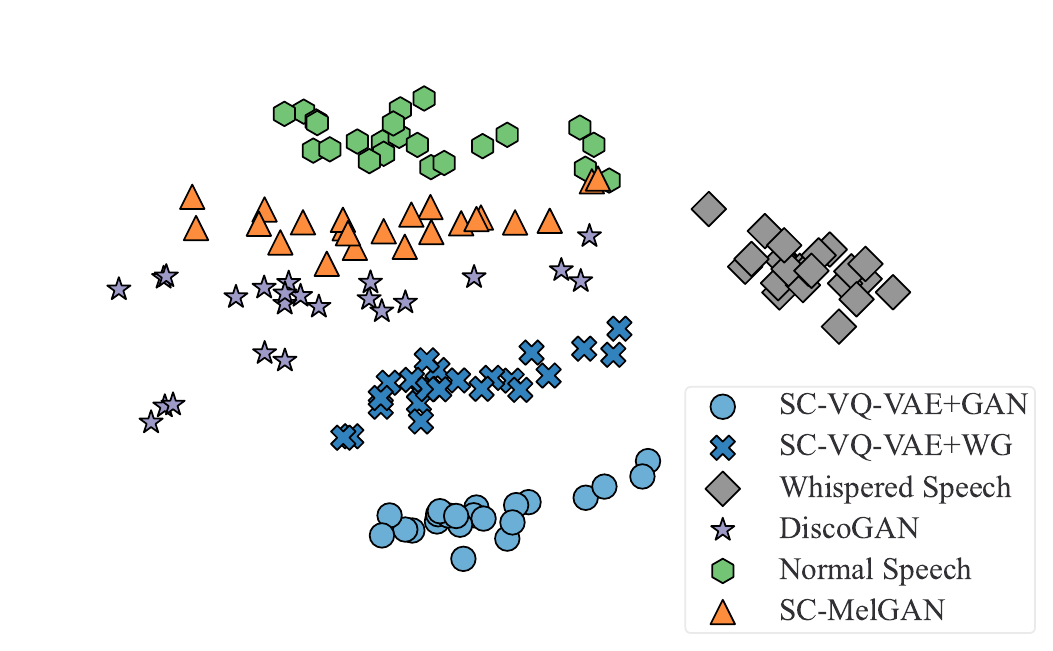}
	\vspace{-8.5mm}
    \caption{Visualization of wav2vec 2.0 speech representations using a t-SNE transform (perplexity = 30).}
  \label{fig:tsne}
\vspace{-3mm}
\end{figure}

We extracted embeddings for each utterance used to conduct the CMOS test in \Cref{ssec:subj} and applied t-distributed stochastic neighbor embedding (t-SNE) \cite{vandermaaten08tsne} to map each datapoint to a location in two-dimensional space. 
\Cref{fig:tsne} shows that embeddings obtained from utterances that have been transformed by one of the four systems, are generally accumulated at locations close to each other. 
Furthermore, embeddings obtained from samples converted by \textit{SC-MelGAN} (triangles) and \textit{DiscoGAN} (stars) are located closest to the embeddings generated from the normal speech samples (hexagons). 
Simultaneously, embeddings computed from samples transformed by \textit{SC-VQ-VAE+WG} (crosses) and \textit{SC-VQ-VAE+GAN} (circles) are located farthest apart from the embeddings representing normal speech samples (hexagons). 

The results in \Cref{fig:tsne} can be confirmed by distance computations in high-dimensional embedding space. 
The mean pairwise cosine distance between embeddings from utterances converted by \textit{SC-MelGAN} and the whispered versions of these utterances is 0.47, while the distance to the normal versions is only 0.17. 
The mean pairwise embedding distances for utterances converted by \textit{DiscoGAN} and the whispered versions of these utterances is 0.46, while the distance to the normal versions is 0.26. 
The embeddings generated for utterances converted by \textit{SC-VQ-VAE+WG} and \textit{SC-VQ-VAE+GAN} are also located closer to those of normal utterances (0.30 and 0.37) than to those of whispered utterances (0.43 and 0.46). 
\vspace{-2mm}
\subsection{Discussion}
\vspace{-1mm}
Our evaluation showed that the \textit{SC-VQ-VAE} architecture performed worse than \textit{SC-MelGAN}. 
Traditional VAEs have the tendency to generate blurry images \cite{larsen16vaegan}. 
Even though VQ-VAEs are less prone to this problem in large-scale image generation tasks, it reappeared in our case. 
The problem becomes visible, when spectrograms of the converted speech are visualized. 
For example, the harmonics of an utterance are clearly visible as horizontal striations in narrowband spectrograms obtained for utterances from \textit{SC-MelGAN} and \textit{DiscoGAN}, but they almost disappear in spectrograms obtained from one of the \textit{SC-VQ-VAE} models. 
We assume that this is mainly due to the limited amount of audio data available. 
Furthermore, the problem of codebook collapse \cite{vandenoord18collapse}, where only a small number of codewords is updated during training, never completely disappeared despite $K$-means initialization. 

The \textit{SC-MelGAN} model outperformed the baseline system throughout all different types of evaluations. 
However, listening tests revealed that in some cases, \textit{SC-MelGAN} introduced artifacts that sound like human speech but don't constitute any intelligible words. 
We assume that this is in parts due to the necessary time alignment for training, which was done using dynamic time warping on the spectral input and target features. 
Therefore, future research will focus on optimizing the alignment process and the removal of the parallel training data requirement. 
\vspace{-2mm}
\section{Conclusions}\label{sec:conclusion}
\vspace{-1mm}
We demonstrate how MelGAN and VQ-VAE systems can be altered to convert whispered speech to normal speech. 
The experiments conducted on speaker-independent models trained on the wTIMIT corpus show that the proposed approaches improve MCD by at least 25\% relative to the \textit{DiscoGAN} baseline. 
Our \textit{SC-MelGAN} yields the largest RMSE improvement (25\% relative to the baseline). 
In the subjective evaluation, the \textit{SC-MelGAN} approach significantly improves naturalness, intelligibility, and voicing, relative to whispered speech.
Additionally, we show that W2V2 embeddings can be a useful tool to illustrate speech conversion results.  
The proposed evaluation method of using pairwise distances between W2V2 embeddings also suggests that transformations obtained with \textit{SC-MelGAN} are closer to normal speech than the baseline. 
\newpage
\bibliographystyle{IEEEbib}
\footnotesize{
\bibliography{strings,refs_fixed}
}

\end{document}